# Floating Electrode Electrowetting on Hydrophobic Dielectric with an SiO$_2$ Layer

*Mehdi Khodayari, Benjamin Hahne, Nathan B. Crane, Alex A. Volinsky*

Department of Mechanical Engineering, University of South Florida, Tampa FL 33620, USA

Floating electrode electrowetting is caused by dc voltage applied to a liquid droplet on the Cytop surface, without electrical connection to the substrate. The effect is caused by the charge separation in the floating electrode. A highly-resistive thermally-grown SiO$_2$ layer underneath the Cytop enables the droplet to hold charges without leakage, which is the key contribution. Electrowetting with an SiO$_2$ layer shows a memory effect, where the wetting angle stays the same after the auxiliary electrode is removed from the droplet in both conventional and floating electrode electrowetting. Floating electrode electrowetting provides an alternative configuration for developing advanced electrowetting-based devices.

Electrowetting is an electromechanical phenomenon [1], in which a small droplet (usually with a volume of nano to micro liters), placed on a hydrophobic dielectric layer or a surface with micro-pillars [2], changes shape upon application of an electric field across the droplet/dielectric substrate. Typically this is quantified in terms of the change in the apparent contact angle. Conventionally, the electric field is created by applying a potential difference between an electrode connected to the droplet and another electrode underneath the dielectric layer [3]. Other configurations are possible, including grounding from below [4-5], bi-directional and continuous electrowetting [6-7]. The wetting angle is given by the Lippmann equation:

$$\cos\theta_1 = \cos\theta_0 + \varepsilon_0\varepsilon_r V^2 / 2\delta\gamma_{LO} \qquad (1)$$

Here, $\theta_0$ and $\theta_1$ are the angles before and after electric field application, $V$ is the applied voltage, $\gamma_{LO}$ is the droplet/second phase surface energy (air in this case), $\delta$ is the dielectric thickness, and $\varepsilon_0\varepsilon_r$ is the dielectric permittivity.

Electrowetting has applications in electrowetting-based screens [8], vibration energy harvesters [9], lenses [10], and lab-on-a-chip devices [11-14]. Electrowetting can be also employed to characterize the formation of crystalline and amorphous phases in droplet bodies. With a new technique, Accardo et. al. have demonstrated the formation of amorphous and crystalline calcium carbonate phases in mixing droplet bodies using the X-ray scattering method [15]. Electrowetting is typically achieved by connecting the substrate electrode and the droplet electrode directly to the power source and ground, respectively. Here, an alternative observation in the electrowetting system is reported, where a liquid droplet is actuated by applying voltage to the droplet placed on top of an isolated silicon wafer. This configuration is referred to as floating electrode electrowetting (FEE). To achieve FEE, the droplet voltage was ramped to both positive and negative values, while the wafer was separated from a grounded stage by a glass slide. For comparison, the conventional electrowetting process was also performed by grounding the silicon wafer below the $SiO_2$ layer. Three different electrolyte solutions, namely 0.1 M NaCl, 0.1 M $Na_2SO_4$, and 0.1 M citric acid were tested, and all behaved similarly. The results obtained with 0.1 M citric acid electrolyte solution are reported here. The experimental setup is shown in Figure 1, with the platinum wire used as the auxiliary electrode.

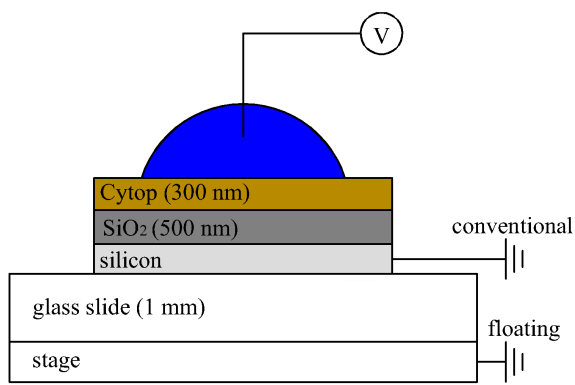

**Figure 1. Schematics of the conventional and floating electrode electrowetting. A platinum wire (0.05 mm in diameter with 99.95% purity) is used as the auxiliary electrode immersed in the droplet.**

In this experiment, the wafers were prepared by thermally growing 500 nm $SiO_2$ layer on the n-type silicon wafers. To make the surface hydrophobic, a 300 nm Cytop layer was spin-coated on top of the oxide layer (pre-baked at 100 °C for 90 sec and then post-baked at 200 °C for 1 h). The droplet profile was imaged digitally, and the contact angle was measured using the ImageJ Drop Analysis plug-in [16]. The results of the contact angle measurements for both conditions are presented in Figure 2.

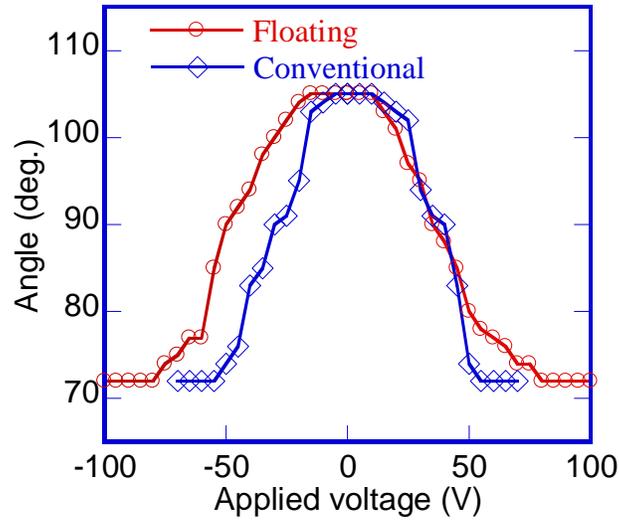

**Figure 2. Contact angle measurements on an oxidized Si wafer coated with 300 nm Cytop in conventional and FEE electrowetting systems. The droplet voltage was ramped to positive and negative values in 5 V increments.**

While the positive voltage curve tracks the conventional case for lower voltages, the negative voltage shows a distinct offset. The FEE droplet saturates at around ±85 V, opposite to ±50 V in conventional electrowetting. Corona charging is not a possible mechanism, since the actuation voltage is far below the voltage required for air ionization [17-18]. Here, FEE is attributed to charge redistribution in the floating electrode. It is proposed that in FEE the charge at the droplet/substrate interface induces a charge separation in the electrode that creates an effective voltage difference across the dielectric. This causes the droplet contact angle modulation. Charge transfer to the droplet from the auxiliary electrode drives the process. This is similar to the results observed by di Virgilio et al [19], except that in this work the charge is applied to the droplet through an electrode rather than by corona charging. Proposed mechanism quantification is outside the scope of this report. The aim of this report is to demonstrate the floating electrode electrowetting process.

The thermally grown $SiO_2$ layer underneath the Cytop provides a highly resistive insulation. To show the high resistance of the thermal $SiO_2$ layer, the conventional electrowetting on a silicon wafer with 500 nm of $SiO_2$ and 300 nm Cytop on top was performed, while the current was measured. The electrical connection was made to the silicon wafer below the $SiO_2$ layer. To compare these results with other materials, the same measurements were performed with aluminum and chromium deposited between the same $SiO_2$ and Cytop layers and the ground connected to the metal electrode. The results are shown in Figure 3 (a) and (b).

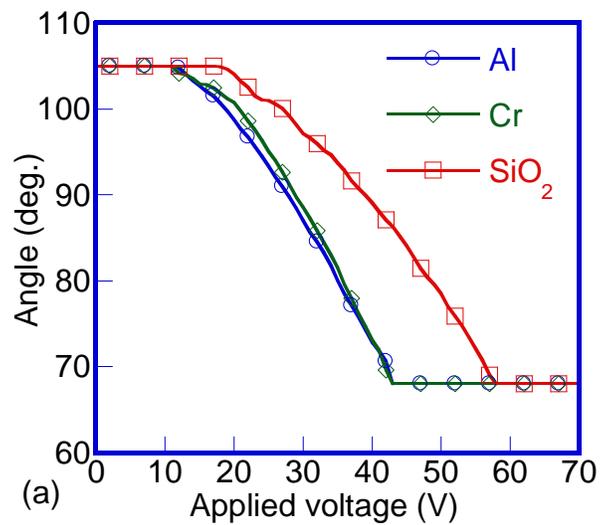
(a)

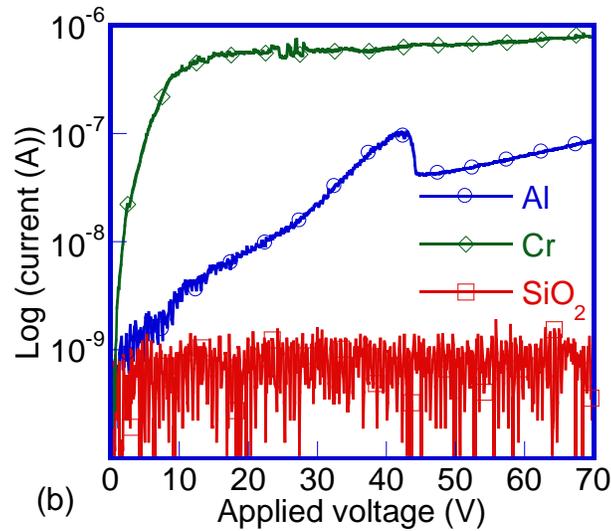

**Figure 3. (a) Contact angle versus voltage and (b) current versus voltage on three different electrowetting substrates, namely Si/SiO$_2$/Cr/Cytop, Si/SiO$_2$/Al/Cytop, and Si/SiO$_2$/Cytop in the conventional electrowetting system (the electrical connections are made to chromium, aluminum, and silicon layers, respectively). In each test, 15 μl droplet of 0.1 M citric acid is placed on the hydrophobic Cytop layer, and then the substrate voltage is ramped up to +70 V in 1 V/70 ms increments with respect to a platinum wire immersed in the droplet. The current passage through the circuit is measured concurrently. To better understand how the droplets contact angle is modulated on the substrates, the contact angle versus voltage curves are also presented in this figure (Figure 3 (a)).**

Figure 3 (a) shows contact angle variation versus voltage on three substrates, namely Si/SiO$_2$/Cr/Cytop, Si/SiO$_2$/Al/Cytop, and Si/SiO$_2$/Cytop (the electrical connections are made to Cr, Al, and Si layer, respectively). Figure 3 (b) shows current versus voltage curves, which indicate a significant difference between the charge transfer resistance of the SiO$_2$/Cytop stack and the single Cytop layer in conventional electrowetting. In each test, a droplet is placed on an electrowetting substrate and the substrate electrical potential is ramped up to +70 V with respect to the droplet (in this test FEE is not performed). With Cytop alone, two different conditions of non-passivating and passivating electrode/electrolyte systems are examined. It is well-known that in passivating systems the charge transfer resistance can be significantly improved [20-21]. However, this test shows that with the $SiO_2$/Cytop dielectric, the charge transfer resistance is even higher than in passivating systems. It will be shown that FEE does not occur with poor dielectric in passivating electrode/electrolyte of passivating systems [21], but it does on the $SiO_2$/Cytop dielectric due to its high charge transfer resistance.

The current magnitude with the chromium layer is the highest, related to the non-passive chromium oxide formation at the Cytop damage sites and the subsequent electrochemical reactions. With aluminum, the current magnitude is less, due to the passive aluminum oxide formation at the damage sites, when aluminum is in contact with citric acid due to the Cytop dielectric local damage [6,20-21]. However, with only an $SiO_2$ layer (no metal layer) the current magnitude remained constant, around 1 nA over the whole voltage ramp, indicating extremely high electrical resistance of the $SiO_2$ layer. This test shows a comparison of the electrical resistance between Cytop alone and the $SiO_2$/Cytop dielectric. $SiO_2$ is a well characterized material, with high resistivity between $10^9$ and $10^{16}$ $\Omega \cdot cm$ [22-23]. In fact, FEE occurs due to the high electrical resistance of the $SiO_2$ layer, which makes the droplet capable of holding charges.

However, when a conductive layer (chromium or aluminum) is deposited between the $SiO_2$ layer and the Cytop, electrowetting does not happen without grounding the electrode, as the Cytop alone cannot provide high enough isolation between the droplet and the conductive layer for the droplet to hold charges. The FEE results with and without the conductive layer are shown in Figure 4(c) and (b), respectively.

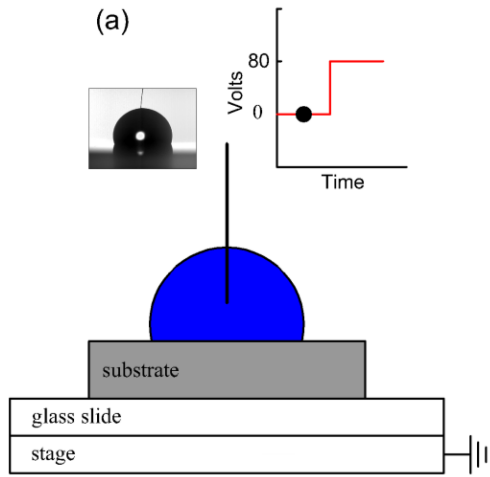

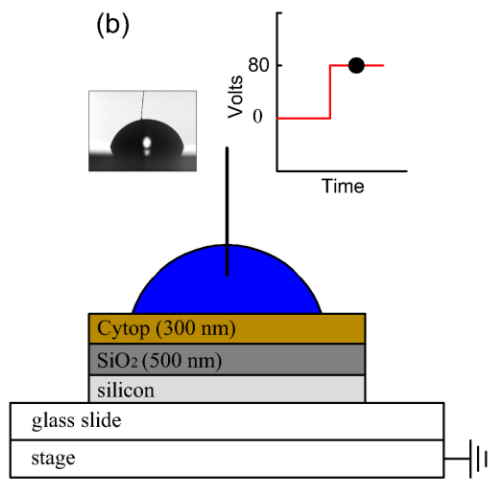

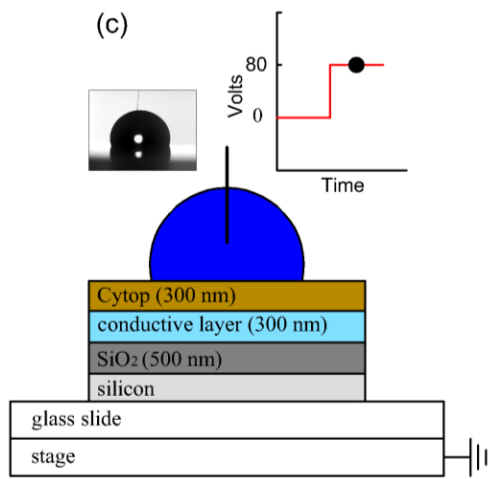

**Figure 4. FEE: (a) on a 500 nm SiO₂ coated with 300 nm Cytop before voltage application; (b) after voltage application without the conductive layer below the Cytop, and (c) with a conductive layer (Cr or Al) between the SiO₂ and the Cytop after voltage application. The insets show the corresponding droplet snapshots. The tests with the same magnitude and opposite polarity of the droplet showed the same results, where FEE occurred only on the wafer without the conductive layer. The solid circles on the applied voltage/time curves show the time at which the droplet snapshots were taken.**

Electrowetting with an SiO₂ layer also exhibits an interesting memory effect. When the platinum electrode was moved from the droplet with an applied voltage, the droplet did not retract to the original wetting angle position, unless a zero voltage was applied and the electrode was reinserted into the droplet. The same effect is seen for FEE and conventional electrowetting with the SiO₂ dielectric on the silicon wafer. In this study, the droplets did not show any retraction over one hour (measurement period) when the power source was turned off after applying 80 V and removing the electrode from the droplet (Figure 5). The droplet volume decreased due to evaporation, but the contact angle stayed the same in both cases. By comparison, an uncharged droplet maintains its initial contact angle during evaporation so that contact angle hysteresis does not influence the measured angle. Readers should also refer to the supplementary video that demonstrates the memory effect observed here (Figure 5 (a)).

A memory effect has been also reported with fluoropolymer dielectric with BaTiO₃ nano-powders due to the charge trapping in the dielectric layer [24]. In this BaTiO₃ system, a reverse voltage is required to change the droplet contact angle to the initial condition, because a residual charge is trapped in the dielectric layer itself. In our experiments, when zero voltage is applied to the droplet, the contact angle switches back to the initial non-wetting value, since the bistability is obtained through trapping of the charges in the droplet. The same effect of the droplet contact angle switching back to the initial non-wetting value is observed when droplet is discharged by grounding it through the top electrode, allowing the charges to escape, as shown in the supplementary video (Figure 5 (a)), with the corresponding timeline in Table 1.

Table 1. Timeline of the floating electrode electrowetting memory effect video in Figure 5 (a).

| Time, sec | Event |
|---|---|
| 0 | Platinum electrode is in the droplet. Silicon electrode is floating. No potential is applied to the platinum auxiliary electrode. |
| 01 | +80 V potential is applied to the platinum electrode. Droplet responds by spreading on the substrate. |
| 05 | Platinum electrode is removed from the droplet with the potential still applied on the platinum electrode. |
| 10 | Platinum electrode potential is set to zero, but it is not in contact with the droplet so the droplet shape is unchanged demonstrating the memory effect. |
| 19 | Platinum electrode with zero potential is inserted into the droplet again. This discharges the droplet and the droplet returns to its initial shape. |

In conclusion, a floating electrode electrowetting can be performed on Cytop-coated thermally oxidized Si substrates. This approach eliminates the need for one of the electrical connections- potentially simplifying the structure of some electrowetting devices. The floating electrode electrowetting with an $SiO_2$ layer also exhibits a memory effect, as the droplet does not show any tendencies to retract to the original wetting angle once the voltage is turned off and the electrode is removed from the droplet.

The authors acknowledge support from the National Science Foundation through CMMI-1130755 grant.

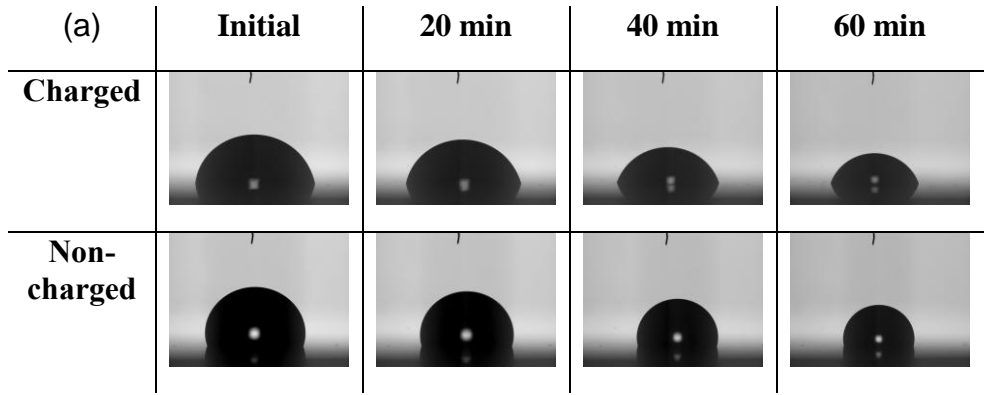

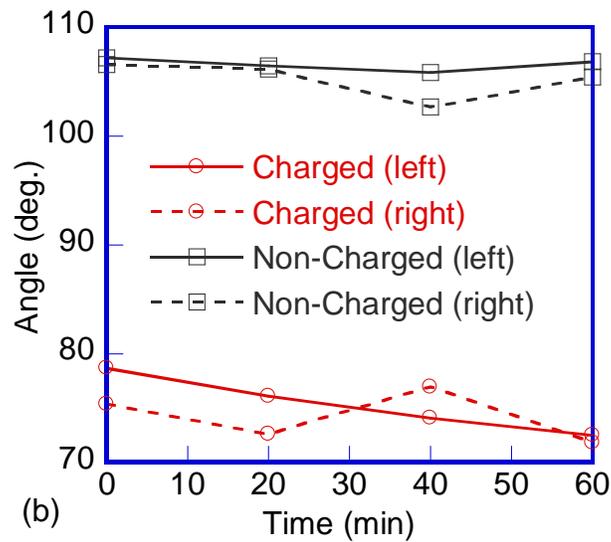

**Figure 5. (a) Memory effect after electrode retrieval in FEE compared with non-charged evaporating droplet. The pictures in the first row show a droplet after applying +80 V in the FEE configuration, and removing the electrode from the droplet (enhanced online). The corresponding video of the droplet charging and discharging steps is available online. The pictures in the second row show a droplet placed on the wafer without charging. In both cases droplet volume decreases due to evaporation, but the contact angle stays the same. (b) Contact angle (left and right angles) of the FEE charged and non-charged droplets.**